\documentclass[12pt]{iopart}
\usepackage[dvipdfm]{graphicx}

\usepackage{bbm}
\begin{document}

\title{Versatile shaper-assisted discretization of energy-time entangled photons}

\author{B Bessire, C Bernhard,  T Feurer and A Stefanov}

\address{Institute of Applied Physics, University of Bern, CH-3012 Bern, Switzerland}

\ead{bbessire@iap.unibe.ch}

\begin{abstract}
We demonstrate the capability to discretize the frequency spectrum of broadband energy-time entangled photons by means of a spatial light modulator to encode qudits in various bases. Exemplarily, we implement three different discretization schemes, namely frequency bins, time bins and Schmidt modes. Entangled qudits up to dimension $d=4$ are then revealed by two-photon interference experiments with visibilities violating a $d$-dimensional Bell inequality. 
\end{abstract}

\maketitle

\section{Introduction}

Entanglement \cite{horodecki2009} is a unique feature of quantum theory having no analogue in classical physics. Spontaneous parametric down-conversion (SPDC) has been used as a source of entangled photon pairs for more than two decades \cite{ghosh1987} and provides an efficient way to generate non-classical states of light for fundamental tests of nature \cite{zeilinger1999,genovese2005}, for quantum information processing \cite{gisin2002,gisin2007,kok2007} or for quantum metrology \cite{giovannetti2011}. Entanglement between two photons emitted by SPDC can occur in one or several (hyperentanglement \cite{barreiro2005}) possible degrees of freedom of light, namely polarization, transverse momentum and energy. Polarization entanglement \cite{kwiat1999} has been used in many experiments even involving multiple pair states \cite{bouwmeester1999}. However, because light supports only two polarization modes, the Hilbert space of each photon is limited to a dimension of two (qubits). On the other hand, entangling $d$-dimensional states denoted as qudits requires multi-mode states of light with $d\geq 2$. This can be for instance achieved using a specific discretization scheme of the transverse momentum degree of freedom. Experiments have been performed where entanglement appears in a discrete set of orbital angular momentum modes \cite{dada2011,pires2009,mair2001,agnew2011,fickler2012,giovannini2012}, pixel modes \cite{sullivan2005} and slit modes \cite{lima2009}. The manipulation and detection of transverse entanglement mainly rely on the ability to experimentally address the transverse momentum modes with the help of holograms or spatial light modulators. The entanglement content is theoretically quantified by the Schmidt number $K$, and is, for usual parameters of the pump laser and the SPDC crystal, on the order of 10 to 50 for transverse wave vector entanglement \cite{law2004,giovannini2012}.

Similar Schmidt numbers are in reach for energy-time entanglement generated by a short pump pulse \cite{law2000,mikhailova2008,brida2009} but considerably higher values of $K$ can be obtained for SPDC driven by a quasi-monochromatic pump laser. While high dimensional entanglement in the transverse momentum modes has been extensively studied, there is still a lack of experiments exploiting the energy degree of freedom to generate qudits for $d>2$. Thus far, qudits with $d=3,4$ were demonstrated via two-photon interferences in \cite{marcikic2002,thew2004,richart2012}. In these experiments, the entanglement was encoded in a time-bin basis realized by interferometers with multiple arms. Time bins have been preferentially used as a basis for entangled qudits because they can be coherently manipulated by interferometers. However, the scalability to higher dimensions becomes prohibitively complex in view of interferometric stability.

By \textit{directly} manipulating the photon spectrum, we demonstrate the implementation of various discretization schemes to realize entangled qudits. For this purpose we coherently address selected spectral components of the entangled photons by a spatial light modulator (SLM) and make use of an ultrafast optical coincidence detection, i.e.~sum-frequency generation (SFG) \cite{peer2005,zaeh2008}. As compared to the aforementioned experiments using interferometric methods, this procedure is intrinsically phase stable and potentially scalable to very high dimensions. It has been used to investigate frequency-bin entangled qudits by quantum state tomography and Bell measurements \cite{bernhard2013}. In this paper, we demonstrate the versatility of the experiment by discretizing the continuous frequency space not only in frequency bins but in different other relevant orthogonal bases. Specifically, in the time-bin basis and a basis obtained by a Schmidt decomposition. Moreover, we show that the presented method is well suited to gain physical insights, e.g.~with respect to the coherence time of the entangled photons in the time-bin basis. The projection onto Schmidt modes further lays the groundwork for a reconstruction of the SPDC state in terms of the Schmidt basis in the frequency domain.   

The paper is organized as follows: In section 2 we introduce the general theoretical framework to discretize a continuous frequency space in a countable subspace together with three specific realizations: Frequency bins, time bins and Schmidt modes. Subsequently, we describe the experimental setup in section 3. In section 4, we experimentally demonstrate and quantify entanglement in the three bases by means of projective measurements equivalent to two-photon interferometry \cite{franson1989} in the time-bin case. Entangled qudits up to $d=4$ are analyzed within the context of a $d$-dimensional Bell inequality. Finally, we conclude this paper by discussing the limitations of the current setup and the possible improvements in view of higher dimensions.

\section{Theory}
\subsection{Theoretical framework}
\begin{figure}[ht]
\begin{center}
\includegraphics{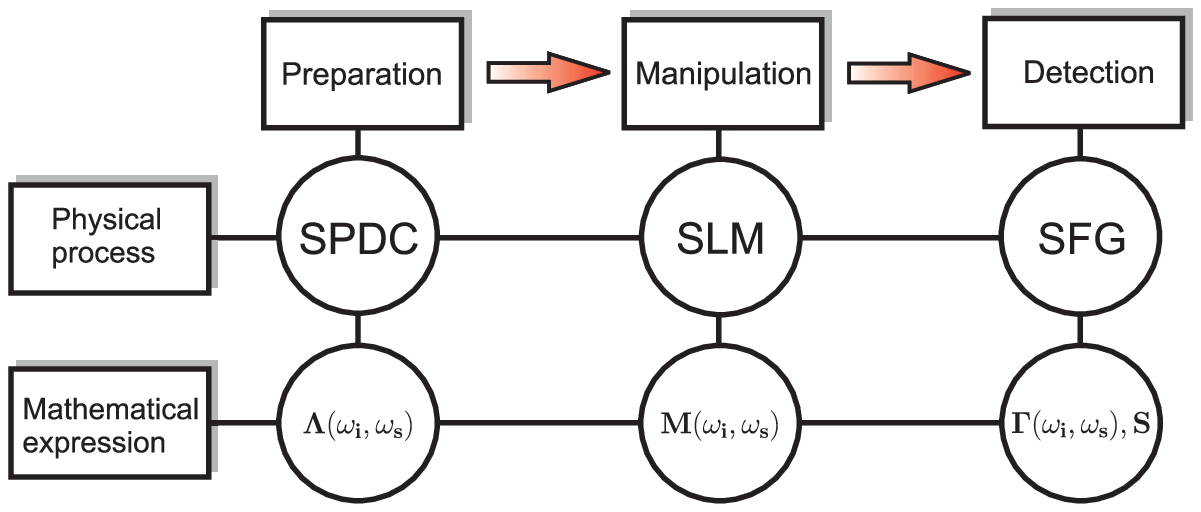} 
\caption{\label{fig:theory_scheme} The three main parts of the experiment: The preparation of energy-time entangled photons through SPDC; the subsequent manipulation of their spectrum using a SLM; and the detection through SFG. Mathematical expressions for $\Lambda(\omega_i,\omega_s)$, $M(\omega_i,\omega_s)$, $\Gamma(\omega_i,\omega_s)$ and $S$ are derived in the corresponding subsections.}  
\end{center}
\end{figure}

In the following, the theoretical framework of the experimental results presented hereafter is discussed according to the schematic shown in figure~\ref{fig:theory_scheme} which allows, in a unified framework, the description of qudits encoding in any energy-time representation. 

\subsubsection{Preparation}
A coherent superposition of energy-time entangled idler ($i$) and signal ($s$) photon pairs occurs through vacuum fluctuations if a pump photon ($p$) is annihilated in a SPDC process. For a configuration where all involved photons are mutually collinear and identically polarized \cite{lerch2013} the corresponding two-photon state reads
\begin{equation}\label{eq:twophstate}
\vert\psi\rangle = \int_{-\infty}^{\infty}\int_{-\infty}^{\infty} d\omega_id\omega_s \Lambda(\omega_i, \omega_s)\hat{a}^{\dagger}_{i}(\omega_i)\hat{a}^{\dagger}_{s}(\omega_s)\vert0\rangle_i\vert0\rangle_s.
\end{equation}
The operators $\hat{a}^{\dagger}_{i,s}(\omega_{i,s})$ act on the combined vacuum state $\mbox{$\vert0\rangle_i\vert0\rangle_s$}$ to create the idler and signal photon with corresponding relative frequencies $\omega_{i,s}=\Omega_{i,s}-\omega_p / 2$. The absolute frequency of the entangled photons is given by $\Omega_{i,s}$ and $\omega_p$ denotes the central frequency of the pump photon. The joint spectral amplitude
\begin{equation}
\Lambda(\omega_i,\omega_s)\propto\alpha(\omega_i,\omega_s)\Phi_{DC}(\omega_i,\omega_s) 
\end{equation}
is written in terms of the pump envelope function $\alpha(\omega_i,\omega_s)$ and the phase matching function $\Phi_{DC}(\omega_i,\omega_s)$ explicitly given by
\begin{equation}
\alpha(\omega_i,\omega_s)=\exp\left(-\frac{(\omega_i+\omega_s)^2\, 2\ln2}{\Delta \omega_p^2}\right),\\
\end{equation}
\begin{eqnarray}\label{eq:phi_DC}
\Phi_{DC}(\omega_i,\omega_s)&= \mathrm{sinc}\left[\frac{\left(\Delta k_{DC}(\omega_i,\omega_s)+\frac{2\pi}{G_{DC}}\right)L_{DC}}{2}\right]\nonumber\\
&\times\exp\left({i\frac{\left(\Delta k_{DC}(\omega_i,\omega_s)+\frac{2\pi}{G_{DC}}\right)L_{DC}}{2}}\right)
\end{eqnarray}
with a pump pulse that has a full width at half maximum of $\Delta \omega_p$ in the spectral intensity. If the nonlinear crystal with length $L_{DC}$ is periodically poled with poling period $G_{DC}$ to achieve quasi-phase matching, then the efficiency for SPDC is optimal if $\Delta k_{DC}\approx -2\pi/G_{DC}$. The phase mismatch $\Delta k_{DC}(\omega_i,\omega_s)=k_i(\omega_i)+k_s(\omega_s)-k_p(\omega_i+\omega_s)$ includes the dispersion properties of the SPDC crystal through its corresponding Sellmeier equations.  

\subsubsection{Manipulation}\label{manipulation}
The spectrum of the entangled photons is manipulated in amplitude and phase by a SLM where the modulator action on each photon is described by a complex transfer function $M^{i,s}(\omega)$. The joint spectral amplitude is transformed by the SLM to
\begin{equation}\label{eq:lambda_mod}
\tilde{\Lambda}(\omega_i,\omega_s)=\Lambda(\omega_i,\omega_s)M(\omega_i,\omega_s)
\end{equation}
with
\begin{equation}\label{eq:M}
M(\omega_i,\omega_s)=M^i(\omega_i)M^s(\omega_s),
\end{equation}
where additional restrictions on $M(\omega_i,\omega_s)$ are time-stationarity and 
\begin{equation}\label{eq:M_condition}
\vert M^{i,s}(\omega)\vert\leq 1.
\end{equation}

\subsubsection{Detection}
In general, coincidence detection is essential to reveal entanglement. The state $\tilde{\Lambda}(\omega_i,\omega_s)$ could be experimentally detected by a combination of narrow-band frequency filters and single photon counters. However, this would yield a signal proportional to $\left|\tilde{\Lambda}(\omega_i,\omega_s)\right|^2$ which is insensitive to any phase modulation in $M(\omega_i,\omega_s)$. To circumvent this problem we seek for a detection scheme that yields a signal which is proportional to $\left|\mathcal{F}\left\{\tilde{\Lambda}(\omega_i,\omega_s)\right\}\right|^2$, the 2D Fourier transform of (\ref{eq:lambda_mod}). Such a scheme requires a time resolution better than the inverse of the photon's spectral bandwidth, which in our experiment is on the order of a few femtoseconds. This is about five orders of magnitudes smaller than the time resolution of the best single photon counters actually available. Therefore, we resort to an optical coincidence method that relies on SFG in a nonlinear crystal. To account for its acceptance bandwidth, we define the modified joint spectral amplitude 
\begin{equation}\label{eq:gamma}
\Gamma(\omega_i,\omega_s)\propto\Lambda(\omega_i,\omega_s)\Phi_{SFG}(\omega_i,\omega_s)
\end{equation}
with
\begin{eqnarray}\label{eq:phi_UC}
\Phi_{SFG}(\omega_i,\omega_s)&= \mathrm{sinc}\left[\frac{\left(\Delta k_{SFG}(\omega_i,\omega_s)-\frac{2\pi}{G_{SFG}}\right)L_{SFG}}{2}\right]\nonumber\\
&\times\exp\left({i\frac{\left(\Delta k_{SFG}(\omega_i,\omega_s)-\frac{2\pi}{G_{SFG}}\right)L_{SFG}}{2}}\right).
\end{eqnarray}
Analogous to $\Phi_{DC}(\omega_i,\omega_s)$, the length and the poling period of the SFG crystal are denoted by $L_{SFG}$ and $G_{SFG}$ with a phase mismatch $\Delta k_{SFG}(\omega_i,\omega_s)=k_p(\omega_i+\omega_s)-k_i(\omega_i)-k_s(\omega_s)$. The temporal resolution of the SFG-based detection process is governed by the inverse width of $\Phi_{SFG}(\omega_i,\omega_s)$ and is sufficiently short. In the following we neglect the additional phase factors in (\ref{eq:phi_DC}) and (\ref{eq:phi_UC}) since they cannot be distinguished from other dispersion contributions in the setup and are assumed to be compensated in the experiment. The detected signal after the SFG process is given by
\begin{equation}\label{eq:signal}
S\propto\left|\int_{-\infty}^{\infty}\int_{-\infty}^{\infty}d\omega_id\omega_s\Gamma(\omega_i,\omega_s)M(\omega_i,\omega_s)\right|^2
\end{equation}
and is sensitive to a phase in the transfer function.

\subsection{Finite spectral resolution}
Because of the finite spectral resolution of the optical setup at the position of the SLM, one given frequency component illuminates several pixels of the SLM. To describe this effect, we convolve $\Gamma(\omega_i, \omega_s)$ with the point spread function
\begin{equation}\label{eq:PSF}
PSF(\omega_i,\omega_s)=\exp\left(-\frac{(\omega_i^2+\omega_s^2)\, 2\ln2}{\Delta \omega_{PSF}^2}\right)
\end{equation}
to obtain 
\begin{equation}\label{eq:gamma_PSF}
\Gamma_{PSF}(\omega_i,\omega_s)\propto(\Gamma\otimes PSF)(\omega_i,\omega_s). 
\end{equation}
The width $\Delta \omega_{PSF}$ depends on the imaging distances and the optical elements within the experimental setup. Equation (\ref{eq:gamma_PSF}) is in particular used to determine the Schmidt basis functions in section \ref{Schmidt}. Figure~\ref{fig:Gamma} depicts $\Gamma(\omega_i,\omega_s)$ and $\Gamma_{PSF}(\omega_i,\omega_s)$ showing that the effect of the PSF is a considerable broadening of the joint spectral amplitude along the diagonal direction.

\subsection{Entanglement quantification}
\begin{figure}[ht]
\includegraphics[width=1\textwidth]{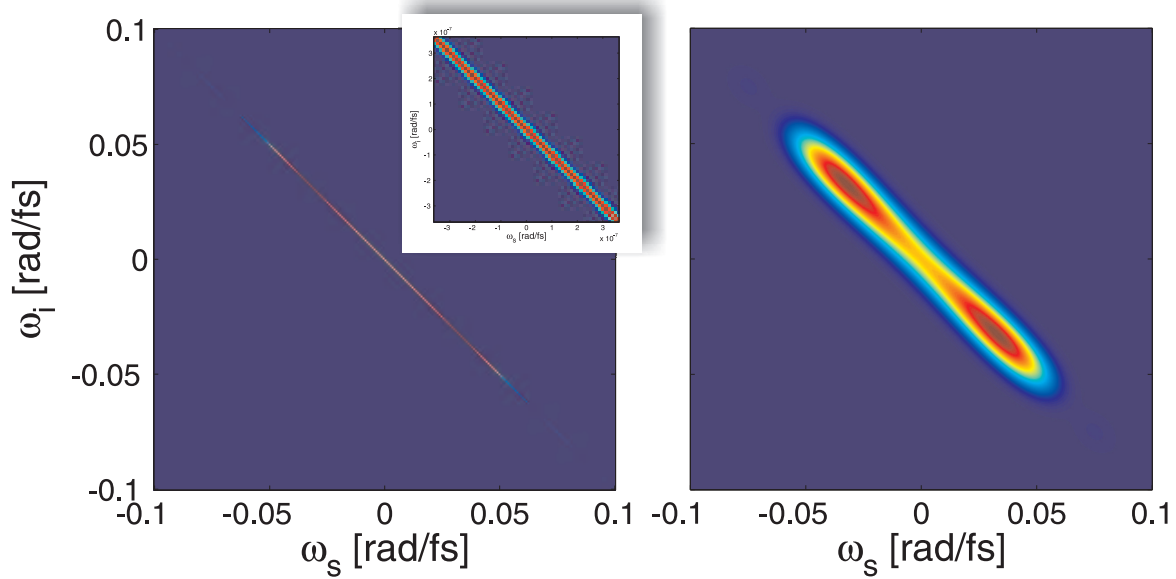} 
\caption{\label{fig:Gamma} Left: $\Gamma(\omega_i,\omega_s)$ for $\Delta \nu_p=5$ MHz, $L_{DC}=L_{SFG}=11.5$ mm and $G_{DC}=G_{SFG}=9$ $\mu$m. Inset: The narrow joint spectral amplitude implies a high degree of entanglement between  idler and signal photon. Right: $\Gamma_{PSF}(\omega_i,\omega_s)$ taking into account the PSF with $\Delta \omega_{PSF}=9.6\times 10^{-3}$ rad/fs.}  
\end{figure}

In order to quantify the degree of entanglement between the idler and signal photon we use the von Neumann entropy $E=-\Tr(\hat{\rho}_{i,s}\log_2 \hat{\rho}_{i,s})$. The entropy is commonly referred to be a valid quantifier of entanglement between two subsystems of a pure entangled state with individual density operators $\hat{\rho}_{i,s}$ \cite{bennett1996}. Through a numerical approximation method \cite{wihler2012} we calculated the entropy of $\Lambda(\omega_i,\omega_s)$ and $\Gamma(\omega_i,\omega_s)$ for a pump spectral bandwidth of $\Delta \nu_p=5$~MHz and further experimental parameters of the preparation and the detection crystals. For $\Lambda(\omega_i,\omega_s)$ we obtain $E=(21.8 \pm 0.1)$~ebits. The entropy is calculated to be $E=(21.1 \pm 0.2)$~ebits using $\Gamma(\omega_i,\omega_s)$ i.e.~we observe almost no influence of the detection process on the degree of entanglement. This amount of entropy is the same as in a maximally bipartite entangled qudit state of dimension $d^2$ with $d=2^{E}\approx 2.2\times 10^6$. This demonstrates that SPDC driven by a spectrally narrow-band pump field offers a potentially very high dimensional state space to encode qudits in frequency modes. Accordingly, we calculate by numerical computation a Schmidt number $K=1/\Tr(\hat{\rho}_{i,s}^2)$ of $K\approx 1.3\times 10^6$. Figure~\ref{fig:Gamma} shows that the effect of the PSF leads to an effective loss in the correlation between the two photons. Consequently, the values for $E$, $d$ and $K$ are reduced to $E \approx 2.6$, as obtained by direct diagonalization of the reduced density matrix, $d \approx 6$ and $K \approx 4.9$.

\subsection{Discretization of the frequency space}
The state (\ref{eq:twophstate}) is a continuous superposition of frequency modes. To encode quantum information in the form of qudits, we project 
\begin{equation}\label{eq:gammatwophstate}
\vert\psi\rangle = \int_{-\infty}^{\infty}\int_{-\infty}^{\infty} d\omega_id\omega_s \Gamma(\omega_i, \omega_s)\hat{a}^{\dagger}_{i}(\omega_i)\hat{a}^{\dagger}_{s}(\omega_s)\vert0\rangle_i\vert0\rangle_s
\end{equation}
into a discrete $d^2$-dimensional subspace spanned by orthonormal product states $\vert j \rangle_i \vert k\rangle_s$ with multi-mode states $\left\vert j\right\rangle_{i,s} \equiv\int_{-\infty}^{\infty}d\omega f^{i,s}_j\left(\omega\right)\hat{a}^{\dagger}_{i,s}(\omega)\left\vert 0\right\rangle_{i,s}$ and $j=0,\ldots,d-1$. The projected state then reads 
\begin{equation}\label{eq:psidisc}
\vert\psi\rangle^{(d)} = \sum_{j=0}^{d-1}\sum_{k=0}^{d-1} c_{jk}\vert j \rangle_i \vert k\rangle_s
\end{equation}
with coefficients 
\begin{equation}
c_{jk}=\int_{-\infty}^{\infty}\int_{-\infty}^{\infty}d\omega_{i}d\omega_{s}f_{j}^{i*}\left(\omega_{i}\right)f_{k}^{s*}\left(\omega_{s}\right)\Gamma\left(\omega_{i},\omega_{s}\right) 
\end{equation}
and the orthonormality condition
\begin{equation}\label{eq:orthonormality}
\int_{-\infty}^{\infty} d\omega f^{i,s \ast}_j(\omega)f^{i,s}_k(\omega)=\delta_{jk}. 
\end{equation}  
Given (\ref{eq:psidisc}), the probability to measure the direct product state
\begin{equation}\label{eq:chi}
\vert\chi\rangle=\left(\sum_{j=0}^{d-1}u^{i*}_j\vert j\rangle_i\right)\left(\sum_{k=0}^{d-1}u^{s*}_{k}\vert k\rangle_s\right) 
\end{equation} 
reads
\begin{equation}\label{eq:projection}
S^{(d)}=\left\vert\langle\chi\vert\psi\rangle^{(d)}\right\vert^2=\left\vert\sum_{j,k=0}^{d-1}u^i_j u^s_k c_{jk}\right\vert^2. 
\end{equation}
If we decompose the transfer function of the SLM into the same basis, i.e.
\begin{equation}\label{eq:mslm}
M^{i,s}(\omega)=\sum_{j=0}^{d-1}u_j^{i,s}f_{j}^{i,s*}(\omega)=\sum_{j=0}^{d-1}|u_j^{i,s}|e^{i\phi_j^{i,s}}f_{j}^{i,s*}(\omega),
\end{equation}
we obtain $S^{(d)}=S$ of (\ref{eq:signal}). Therefore, the measured signal $S$ is given by the projection of the state $\vert\psi\rangle^{(d)}$ onto $\vert\chi\rangle$, and the SLM together with a SFG coincidence detection performs a projective measurement. Because $|u_j^{i,s}|$ and $\phi_j^{i,s}$ can be adjusted independently, any state $\vert\chi\rangle$ can be implemented provided the conditions in section~\ref{manipulation} are fulfilled. 

Through a judicious choice of $f_{j}^{i,s}(\omega)$, various discretization schemes can be realized with the SLM using (\ref{eq:mslm}). Here, we present three different basis functions $f_{j}^{i,s}(\omega)$ to encode qudits in the frequency domain.

\subsubsection{Frequency-bin basis}
\begin{figure}[ht]
\begin{center}
\includegraphics[width=0.5\textwidth]{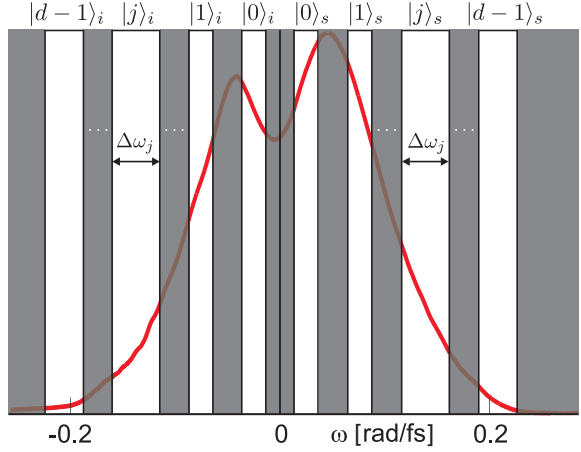} 
\caption{\label{fig:freq_bins}Measured SPDC spectrum overlaid with a schematic frequency-bin pattern. The transmitted amplitude $\vert u_j^{i,s}\vert$ (white bars) of each bin can be adjusted through amplitude modulation by means of the SLM.}  
\end{center}
\end{figure}

An intuitive method to discretize the frequency space is to subdivide the spectrum into frequency bins through amplitude modulation (figure~\ref{fig:freq_bins}). 
The corresponding $f_{j}^{i,s}(\omega)$ are defined according to
\begin{equation}\label{eq:freq_bins}
f_{j}^{i,s}(\omega) = 
\cases{
  1/\sqrt{\Delta\omega_j} & for $\vert\omega-\omega_{j}\vert<\Delta\omega_j/2$ \cr
  0 & otherwise, \cr
}
\end{equation}
where we impose $|\omega_j-\omega_k|>(\Delta\omega_j+\Delta\omega_k)/2$ for all $j,k$ to guarantee that adjacent bins do not overlap. If we further assume a continuous wave pump, we restrict (\ref{eq:psidisc}) to its diagonal form
\begin{equation}\label{eq:psidiscdiag}
\vert\psi\rangle^{(d)} = \sum_{j=0}^{d-1} c_{j}\vert j \rangle_i \vert j\rangle_s.
\end{equation}

\subsubsection{Time-bin basis}
\begin{figure}[ht]
\includegraphics[width=1\textwidth]{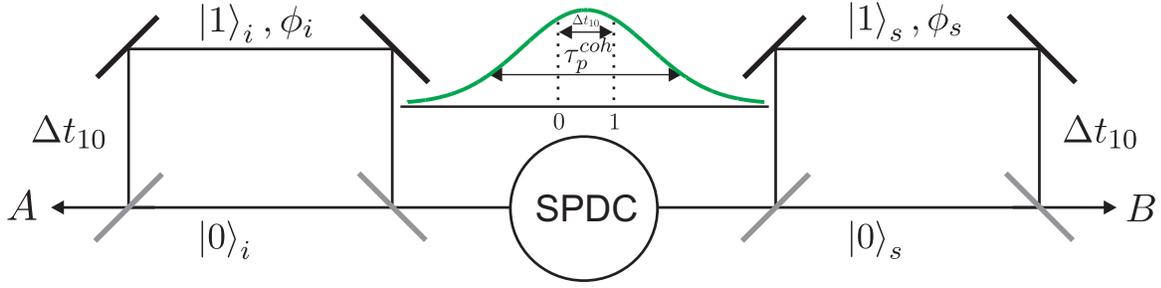} 
\caption{\label{fig:time_bins_franson} Scheme to analyze energy-time entangled two-photon states proposed by Franson \cite{franson1989}. A photon pair is generated by SPDC within the coherence time $\tau_p^{coh}$ of a pump photon. Both photons are injected into two separated and unbalanced Mach-Zehnder interferometers with a time delay of $\Delta t_{10}$ imprinted on the photon in the longer arm. A qubit state is then measured by varying the phases $\phi_i$ and $\phi_s$ while performing coincidence measurements between output ports $A$ and $B$.}  
\end{figure}

Time-bin entangled photons are typically manipulated in the temporal domain by interferometers with variable optical path lengths \cite{marcikic2002,thew2004,richart2012}. Up to now, time bins have been preferentially used to encode quantum information and to analyze the temporal properties of the down-converted photons, a concept that was first proposed by Franson \cite{franson1989}. A Franson interferometer was imitated by shaping the spectrum of entangled photons in the telecom wavelength regime using a wave shaper in combination with conventional electronic coincidence counting in \cite{lukens2013}. In Franson's interferometric scheme, each photon of an entangled pair enters an unbalanced Mach-Zehnder interferometer where both photons undergo the same time delay $\Delta t_{10}$ when traveling along the long path (figure~\ref{fig:time_bins_franson}).

\begin{figure}[ht]
\includegraphics[width=1\textwidth]{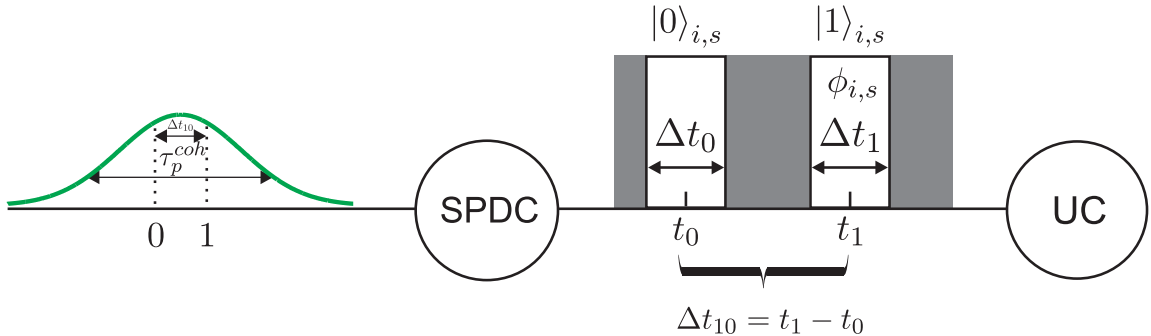} 
\caption{\label{fig:time_bins} Franson's original scheme adapted to time bins implemented by a SLM. Under specific conditions for $\Delta t_{10}$ and $\Delta t_j$ (see text) projections onto a superposition of $\vert 0 \rangle_i \vert 0\rangle_s$ and $\vert 1 \rangle_i \vert 1\rangle_s$ can be measured.}  
\end{figure}

For $\Delta t_{10}\ll\tau^{coh}_p$, where $\tau^{coh}_p$ denotes the coherence time of the pump photon, the state generated by the first pair of beam splitters is a coherent superposition 
\begin{equation}\label{eq:psidisc_qubit}
\vert\psi\rangle^{(2)} = \sum_{j=0}^{1}\sum_{k=0}^{1} c_{jk}\vert j \rangle_i \vert k\rangle_s,
\end{equation}
where we associate $\vert 0\rangle_{i,s}$ with the short and $\vert 1\rangle_{i,s}$ with the long path of the interferometer. Characteristic qubit interference fringes can be observed by a coincidence detection between the output ports $A$ and $B$ while varying the phases $\phi_i$ and $\phi_s$. The original experiment of Franson can be extended to higher dimensional qudits by endowing each interferometer with additional arms. Two four-arm interferometers were used in \cite{richart2012} to experimentally demonstrate energy-time entangled ququarts. Instead of using interferometers, we discretize the time domain of the idler and signal photons into time bins (figure~\ref{fig:time_bins}) 
\begin{equation}\label{eq:time_bins}
f_{j}^{i,s}(t) = 
\cases{
  \sqrt{\frac{2\pi}{\Delta t_j}} & for $\vert t-t_{j}\vert<\Delta t_j/2$ \cr
  0 & otherwise \cr
}
\end{equation}
analogous to (\ref{eq:freq_bins}) in the frequency domain with the help of the SLM. 
The coefficients in (\ref{eq:psidisc_qubit}) are now related to the joint temporal amplitude $\Upsilon(t_i,t_s)$ of the SPDC photons through
\begin{equation}
c_{jk}=\int_{t_j-\frac{\Delta t_j}{2}}^{t_j+\frac{\Delta t_j}{2}}\int_{t_k-\frac{\Delta t_k}{2}}^{t_k+\frac{\Delta t_k}{2}}dt_{i}dt_{s}\Upsilon\left(t_{i},t_{s}\right),
\end{equation}
where $\Upsilon\left(t_{i},t_{s}\right)$ is the Fourier transform of $\Gamma(\omega_i,\omega_s)$. Since the SLM manipulates the frequency spectrum of the entangled photons, the time bins of (\ref{eq:time_bins}) are Fourier transformed to obtain
\begin{equation}\label{eq:time_bins_freq}
f_{j}^{i,s}(\omega) = \sqrt{\frac{\Delta t_j}{2\pi}}e^{-i\omega t_j}\mathrm{sinc}\left(\frac{\omega \Delta t_j}{2}\right),
\end{equation}
where orthonormality holds for any $j$ and $k$ provided that $|t_j-t_k|>(\Delta t_j+\Delta t_k)/2$. In order to attain entangled qubits, the time delay $\Delta t_{10}=t_1-t_0$ has to exceed the coherence time $\tau^{coh}_{i,s}$ of the idler and signal photon to avoid single photon interference. A coincidence window of SFG detection is typically on the order of a few femtoseconds, which itself corresponds to $\tau^{coh}_{i,s}$ for broadband SPDC emission. It is guaranteed for $\Delta t_{10}>\tau^{coh}_{i,s}$ that no $\vert 0 \rangle_i \vert 1\rangle_s$ and $\vert 1 \rangle_i \vert 0\rangle_s$ events contribute to the coincidence signal. Further, to prevent (\ref{eq:time_bins_freq}) acting as a filter on the entangled photons spectrum, $\Delta t_j$ is restricted to $\Delta t_j\ll\tau^{coh}_{i,s}$. Given the constraints on $\Delta t_{10}$ and $\Delta t_j$ are satisfied, the event that both photons of a pair, down-converted at a given time, pass time bin $f_1^{i,s}(t)$ (long arm) cannot be distinguished from two photons created after a time delay $\Delta t_{10}$ and traveling through bin $f_0^{i,s}(t)$ (short arm). The state generated by two time bins then reads
\begin{equation}\label{eq:psidisc_qubit_entangled}
\vert\psi\rangle^{(2)} = c_{0}\vert 0 \rangle_i \vert 0\rangle_s+c_{1}\vert 1 \rangle_i \vert 1\rangle_s.
\end{equation} 
To demonstrate the equivalence between Fransons (FR) interferometric scheme and the implementation of time bins in the frequency space, we consider the transfer function for a single photon in a Mach-Zehnder interferometer
\begin{equation}\label{eq:M_franson}
M_{FR}^{i,s}(\omega)=T+R\,\e^{i\bar{\phi}_{i,s}(\omega)},
\end{equation}
where both beam splitters have transmission and reflection coefficients $T$ and $R$. The total phase shift $\bar{\phi}_{i,s}(\omega)=\omega\Delta t_{10}+\phi_{i,s}$ is the sum of the phase difference $\omega\Delta t_{10}$ between the long and the short arm and an additional absolute phase $\phi_{i,s}$. The total transfer function of the scheme depicted in figure~\ref{fig:time_bins_franson} then reads $M_{FR}(\omega_i,\omega_s)=M_{FR}^{i}(\omega_i)M_{FR}^{s}(\omega_s)$. If we restrict in (\ref{eq:time_bins_freq}) the width of the bins to $\Delta t_0=\Delta t_1=\Delta t$ and put $t_0=0$~fs such that $\Delta t_{10}=t_1$, the general transfer function from (\ref{eq:mslm}) transforms to
\begin{equation}\label{eq:M_time_bins}
M^{i,s}(\omega)=\sqrt{\frac{\Delta t}{2\pi}}\mathrm{sinc}\left(\frac{\omega \Delta t}{2}\right)\left(\vert u_0^{i,s}\vert+\vert u_1^{i,s}\vert\e^{i\bar{\phi}_{i,s}(\omega)}\right)
\end{equation}
for $d=2$ with $\bar{\phi}_{i,s}(\omega)=\omega t_1+\phi_{i,s}$. In the limit $\Delta t_j\rightarrow 0$, the time bins in (\ref{eq:time_bins}) are reduced to $f_{j}^{i,s}(t)=\sqrt{2\pi}\delta(t-t_j)$. Equation (\ref{eq:M_time_bins}) then takes the form 
\begin{equation}\label{eq:time_bin_dt0}
M^{i,s}(\omega)=\frac{1}{\sqrt{2\pi}}\left(\vert u_0^{i,s}\vert+\vert u_1^{i,s}\vert\,\e^{i\bar{\phi}_{i,s}(\omega)}\right)
\end{equation}
and is equal to (\ref{eq:M_franson}) with $\vert u_0^{i,s}\vert=T$ and $\vert u_1^{i,s}\vert=R$ up to a normalization constant.

\subsubsection{Schmidt mode basis}
\label{Schmidt}
\begin{figure}[ht]\label{fig:schmidt_modes}
\begin{center}
\includegraphics[width=0.7\textwidth]{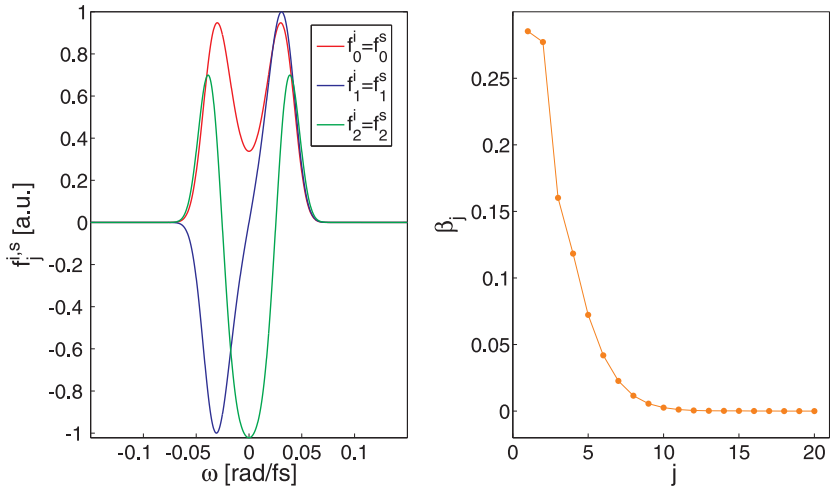} 
\caption{\label{fig:schmidt_functions} Left: Exemplary, the first three Schmidt modes $f^{i,s}_j\left(\omega\right)$ of (\ref{eq:gamma_schmidt}). Right: Shown are the eigenvalues $\beta_j$ up to $j=20$.}  
\end{center}
\end{figure}

The Schmidt decomposition for continuous variable systems and its application to quantify entanglement has been extensively studied in \cite{law2000,parker2000}. To decompose a state which is as close as possible to the state at the position of the SLM, we have to consider the modified joint spectral amplitude $\Gamma_{PSF}(\omega_i,\omega_s)$ taking into account the effect of the finite spectral resolution. Because of being bipartite and pure, the two-photon state of (\ref{eq:gammatwophstate}) can be represented in a Schmidt decomposition
\begin{eqnarray}
\Gamma_{PSF}\left(\omega_{i},\omega_{s}\right)&=\sum_{j=0}^{\infty}\sqrt{\beta_j}\,f^{i}_j\left(\omega_i\right)f^{s}_j\left(\omega_s\right)\nonumber\\
&\approx\sum_{j=0}^{d-1}\sqrt{\beta_j}\,f^{i}_j\left(\omega_i\right)f^{s}_j\left(\omega_s\right),\label{eq:gamma_schmidt}
\end{eqnarray}
where the real valued functions $f^{i,s}_j\left(\omega\right)$ are the eigenvectors or Schmidt modes of the reduced density operators and $\beta_j$ the corresponding eigenvalues. The Schmidt modes itself are orthogonal and form a complete basis. As can be seen in figure~\ref{fig:schmidt_functions}, only a few $\beta_j$ are significantly nonzero if we decompose $\Gamma_{PSF}\left(\omega_{i},\omega_{s}\right)$ with the parameters of our experimental setup. By substituting $\Gamma_{PSF}\left(\omega_{i},\omega_{s}\right)$ in (\ref{eq:gammatwophstate}) and using (\ref{eq:gamma_schmidt}) one finds
\begin{equation}\label{eq:qudit_schmidt}
\vert\psi\rangle\rightarrow\vert\psi\rangle^{(d)}=\sum_{j=0}^{d-1}c_j\,\vert j \rangle_i \vert j\rangle_s
\end{equation}
with $c_j=\sqrt{\beta_j}$. The Schmidt basis thus provides a direct way to discretize the state (\ref{eq:twophstate}) into an entangled qudit state whose dimensionality is only bound by the number of nonzero $\beta_j$. For a symmetric joint spectral amplitude $\Gamma_{PSF}(\omega_i,\omega_s)$ (figure~\ref{fig:Gamma}) we find $f^{i}_j\left(\omega\right)=f^{s}_j\left(\omega\right)$. The Schmidt decomposition has been performed numerically by discretizing the continuous function $\Gamma_{PSF}(\omega_i,\omega_s)$ on a lattice with size $2049 \times 2049$. 

\section{Experimental setup}
\begin{figure}[ht]
\includegraphics[width=1\textwidth]{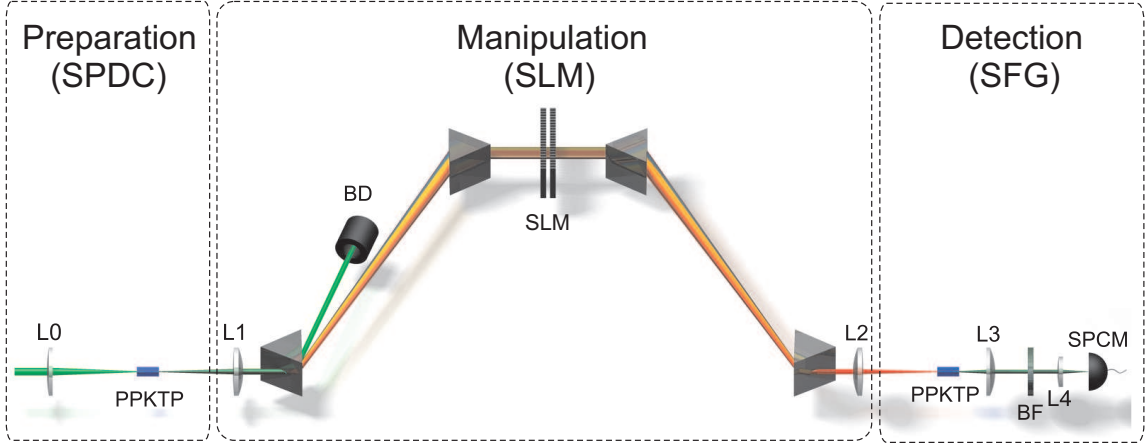} 
\caption{\label{fig:Setup} Schematic of the experimental setup. Preparation: L0 pump beam focusing lens ($f = 150$~mm), PPKTP nonlinear crystal for SPDC. Manipulation: BD beam dump, SLM spatial light modulator, symmetric two lens (L1, L2) imaging arrangement ($f = 100$~mm) to magnify the spectral resolution by 1:6, four-prism compressor. Detection: PPKTP nonlinear crystal for SFG, BF bandpass filter, SPCM single photon counting module with a two lens (L3, L4) imaging system.}  
\end{figure}

The experimental setup shown in figure~\ref{fig:Setup} is composed of three parts: The entangled state preparation by SPDC, the spectral manipulation with the SLM and the coincidence detection by SFG.
Energy-time entangled photons are created by SPDC in a periodically poled KTiOPO$_4$ (PPKTP) crystal with length $L_{DC}=11.5$~mm and a poling periodicity of $G_{DC}=9$ $\mu$m. The pump is a single mode 5~W Nd:YVO$_{4}$ (Coherent Verdi V5) laser operating at 532~nm with a spectral bandwidth of about $\Delta \nu_p=5$~MHz. According to type-0 phase matching, the created idler and signal photons have the same polarization as the pump photon. The operating temperature of the PPKTP crystal is optimized for almost degenerate and collinear emission with a spectral width of the down-converted photons of $\Delta \lambda_{DC}\approx 105$~nm centered around 1064~nm.  

To control the SPDC spectrum, it is dispersed in a symmetric four-prism compressor consisting of equilateral N-SF11 prisms in minimum deviation geometry. At the same time, the prism compressor serves to compensate for the total accumulated group velocity dispersion in the setup and to deflect the residue of the pump into a beam dump. At the symmetry axis of the prism compressor the dispersed spectrum passes two identical nematic liquid crystal arrays of a programmable SLM (Jenoptik, SLM-S640d). Both arrays consist of 640 pixels each 100~$\mu$m wide and separated by a gap of 3~$\mu$m from its nearest neighbors. For a given set of basis functions $f_j^{i,s}(\omega)$, the transmitted frequencies at each pixel are manipulated independently in amplitude and phase according to the transfer function (\ref{eq:mslm}) by adjusting the orientation of the nematic molecules with a specific voltage.

A second, identical, PPKTP crystal is phase matched to detect entangled photons in coincidences through SFG \cite{peer2005}. This provides a coincidence time window with femtosecond temporal resolution. The sum-frequency photons are detected by a single photon counting module (ID Quantique, id100-50-uln) and the remaining IR photons are filtered by a bandpass filter (4~mm BG18). According to reference \cite{dayan2005}, the maximal allowed flux of down-converted photons at the single photon limit is given by $\Phi_{max}\approx\Delta\nu_{DC}$. A spectral bandwidth of $\Delta \lambda_{DC}\approx 105$~nm corresponds to a maximal flux of $\Phi_{max}=2.8\times 10^{13}$ photons per second or a maximal power of $P_{max}=5.2$~$\mu$W. That is, for the actual power of 1~$\mu$W we find a spectral mode density of $n=P/P_{max}=0.2$ which assures that we are below the single photon limit and therefore no coincidences are measured between photons of different pairs.

\section{Experimental results}

\subsection{CGLMP inequality}\label{sec:cglmp}
We assume the qudits in our experiment to be described by a symmetric noise model
\begin{equation}\label{eq:snmodel}
\hat{\rho}^{(d)}=\lambda_d\vert\psi\rangle^{(d)}\,^{(d)}\langle\psi\vert+\frac{(1-\lambda_d)}{d^2}\mathbbm{1}_{d^2},
\end{equation}
where deviations from a pure state are quantified by the mixing parameter $\lambda_d$ and $\mathbbm{1}_{d^2}$ denotes the $d^2$-dimensional identity operator. Here, $\vert\psi\rangle^{(d)}$ is a maximally entangled state
\begin{equation}\label{eq:qudit_maxent_phase}
\vert\psi\rangle^{(d)}=\frac{1}{\sqrt{d}}\sum_{l=0}^{d-1}\,\e^{i l \phi_0}\vert l \rangle_i \vert l\rangle_s,
\end{equation}
where $\phi_0$ accounts for small phase shifts due to non-perfectly compensated dispersion in the setup. In order to show entanglement without performing full quantum state tomography we make use of Bell test measurements. Collins \textit{et al.} (hereafter referred to as CGLMP) introduced a dimensional dependent Bell parameter $I_d$ to study the non-classical correlations of $d$-dimensional bipartite quantum states in the context of a new family of Bell inequalities \cite{collins2002}. The parameter $I_d$ is related to its corresponding Bell operator $\hat{B}$ \cite{braunstein1992} according to $I_d=\Tr\left(\hat{B}\hat{\rho}^{(d)}\right)$ where it was shown in \cite{terhal2000} that $\hat{B}$ itself can be used as an entanglement witness. Consequently, the violation of a Bell inequality, i.e.~$I_d>2$, indicates entanglement between the two systems involved. Due to a left-open locality loophole in our detection method, however, the subsequent experimental results can not be considered as a test of non-locality. Because of $\Tr\left(\hat{B}\hat{\rho}^{(d)}\right)=\lambda_d I_d^{max}$, the aforementioned inequality can be reformulated in terms of the mixing parameter
\begin{equation}
\lambda_d>\frac{2}{I_d^{max}}\doteq\lambda_d^c
\end{equation}
with its critical value $\lambda_d^c$. The value of $I_d$ is related to the visibility of two-photon interferences which are obtained by projecting (\ref{eq:snmodel}) onto 
\begin{equation}\label{eq:chi_phi}
\vert\chi\rangle=\frac{1}{d}\left(\sum_{j=0}^{d-1}\e^{-i j \phi_i}\vert j \rangle_i\right)\left(\sum_{k=0}^{d-1}\e^{-i k \phi_s} \vert k\rangle_s\right)
\end{equation} 
with $\phi_i=\phi_s=\phi$. The phase $\phi$ is varied by the SLM and controls the interference in the following experiments. The theoretical coincidence signals then read 
\begin{eqnarray}\label{eq:fitd2}
S_\lambda ^{(2)}(\phi)&=\Tr\left(\hat{\rho}^{(2)}\vert\chi\rangle\langle\chi\vert\right)\nonumber\\
&\propto 1+\lambda_2\cos(2\phi+\phi_0),
\end{eqnarray}
\begin{eqnarray}\label{eq:fitd3}
S_\lambda ^{(3)}(\phi)&=\Tr\left(\hat{\rho}^{(3)}\vert\chi\rangle\langle\chi\vert\right)\nonumber\\
&\propto 3+2\lambda_3\left[2\cos(2\phi+\phi_0)+\cos (2(2\phi+\phi_0))\right],
\end{eqnarray}
\begin{eqnarray}\label{eq:fitd4}
S_\lambda ^{(4)}(\phi)&=\Tr\left(\hat{\rho}^{(4)}\vert\chi\rangle\langle\chi\vert\right)\nonumber\\
&\propto 4+2\lambda_4[3\cos(2\phi+\phi_0)+2\cos(2(2\phi+\phi_0))\nonumber\\
&\quad+\cos (3(2\phi+\phi_0))]
\end{eqnarray}
and are used to fit the experimental data with the free parameters $\phi_0$ and $\lambda_d$. The latter accounts for white noise as well as for the point spread function and for couplings between frequency and transverse modes due to a possible misalignment in the experimental setup.  Since we experimentally demonstrate qudits in terms of interference patterns, we make use of the fact that $\lambda_d$ can be related to a visibility \cite{thew2004} according to
\begin{equation}\label{eq:cglmp_lamb}
V_d=\frac{d\lambda_d}{2+\lambda_d (d-2)}.
\end{equation}
Entanglement between idler and signal photon is then present if 
\begin{equation}\label{eq:cglmp_vis}
V_d>V_d(\lambda_d^c)\doteq V_d^c. 
\end{equation}
The values for the critical visibility $V_d^c$ are listed in table~\ref{tab:freq_bin}. To relate CGLMP's inequality to the visibility of interference fringes has the advantage that a possible phase shift $\phi_0$ present in the experiments has no influence on the violation of a Bell inequality. This would be the case if single projection measurements at fixed phase settings were used to determine the value of a Bell parameter.
 
\subsection{Frequency-bin basis}
\begin{figure}[ht]
\begin{center}
\includegraphics[width=1\textwidth]{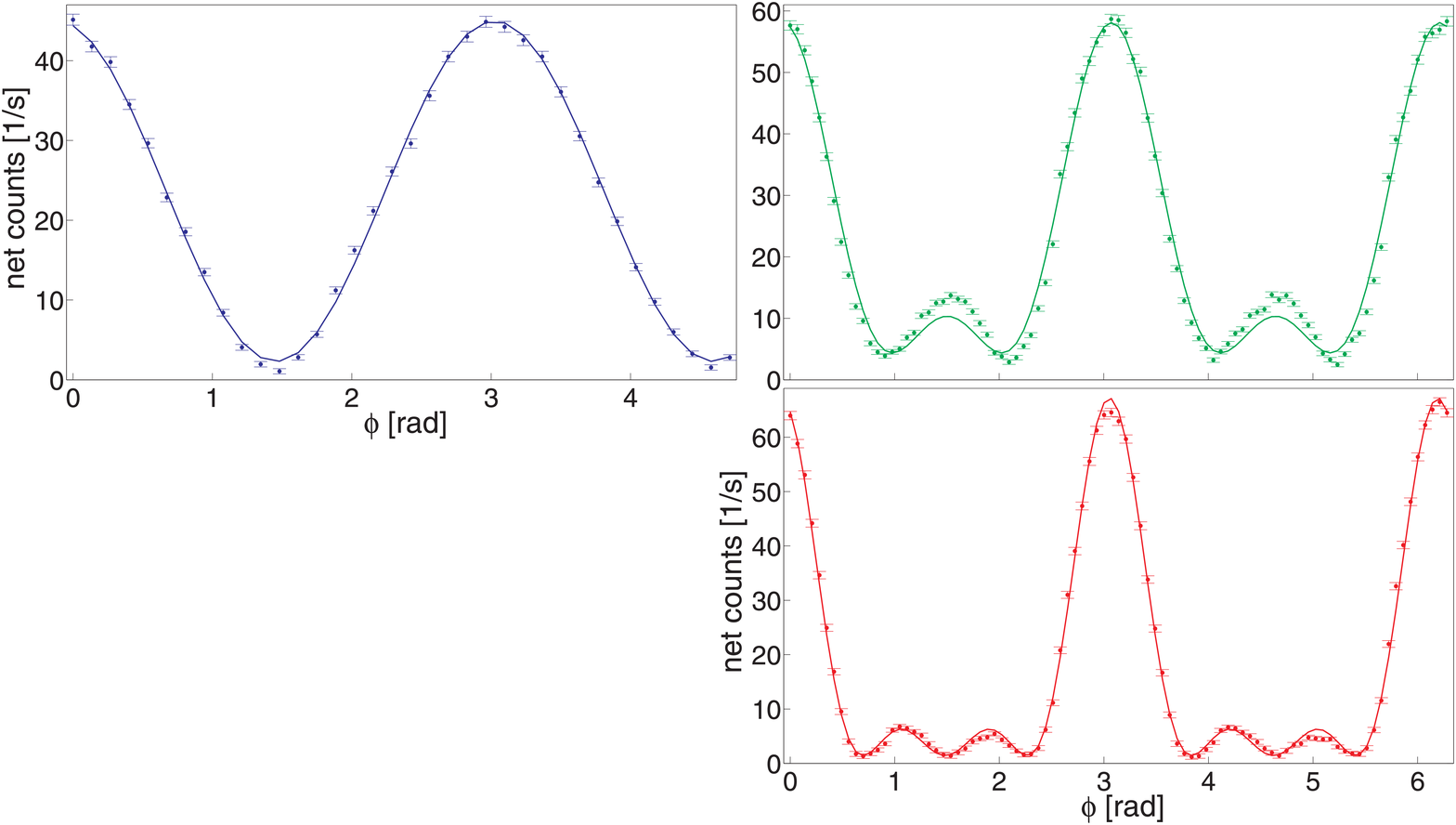} 
\caption{\label{fig:freq_bin_qudits} Frequency-bin basis: Two-photon interferences for a maximally entangled qubit (blue, top left), qutrit (green, top right) and ququart (red, bottom). Shown are background-subtracted coincidence counts (net counts) with 1$\sigma$ standard deviations using Poisson statistics. The solid curves are fits to the data points by means of (\ref{eq:fitd2}), (\ref{eq:fitd3}) and (\ref{eq:fitd4}).}
\end{center}
\end{figure}

With the goal to maximize the entanglement in
\begin{equation}
\vert\psi\rangle^{(d)} = \sum_{j=0}^{d-1} c_{j}\vert j \rangle_i \vert j\rangle_s
\end{equation}
we make use of the Procrustean method of entanglement concentration \cite{bennett1996}. In general, this method equalizes the amplitudes in a partially entangled state through local operations where contributions with higher probabilities are diminished by appropriate filtering. One is then left with a maximally entangled state according to (\ref{eq:qudit_maxent_phase}) with all amplitudes being equal. For a bin structure according to (\ref{eq:freq_bins}) we experimentally equate the $\vert c_j\vert$ by performing single projection measurements onto $\vert\chi_k\rangle=\vert u_k\vert^2\vert  k\rangle_i\vert k\rangle_s$ for $k=0,\ldots, d-1$ with corresponding coincidence signals $S_k=\left\vert\vert u_k\vert^2 \vert c_k\vert\right\vert^2$. The amplitude of the $k$-th frequency bin $\vert u_k\vert\doteq\vert u^i_k\vert=\vert u^s_k\vert$ is then adjusted with the SLM such that all $S_k$ are equal to $S_{min}=\min\limits_{k=0,\ldots, d-1}\{S_k\}$ within the statistical uncertainties. To minimize the photon loss, the Procrustean filtering is accompanied with an optimization of the bin widths $\Delta \omega_j$ where bins at the far end of the spectrum are chosen to be wider than bins located at the center of the spectrum. The measurement time for each $S_k$ was 300~s with a SPCM background coincidence rate of about 11~Hz. Two-photon interference fringes were then measured by a projection onto the state (\ref{eq:chi_phi}) with the SLM using the transfer function (\ref{eq:mslm}) corresponding to frequency bins and scanning the phase parameter $\phi$ in discrete steps while detecting coincidence counts. Note, that $j \phi$ is the resulting absolute phase associated with bin $j=0,\ldots, d-1$ on the idler side of the spectrum. Analogue, we have $k \phi$ for bin $k=0,\ldots, d-1$ in the spectral domain of the signal photon. Figure~\ref{fig:freq_bin_qudits} depicts the measured two-photon interference curves for qudits up to $d=4$.
Equations~(\ref{eq:fitd2}), (\ref{eq:fitd3}) and (\ref{eq:fitd4}) are used to fit the data points. The corresponding visibilities (table~\ref{tab:freq_bin}) are calculated with the aid of the fitting parameter $\lambda_d$ and the relation (\ref{eq:cglmp_lamb}). We find $V_d>V_d^c$ for all $d$ which demonstrates the existence of frequency-bin entanglement. 

\begin{table}
\caption{\label{tab:freq_bin}Frequency-bin basis: Critical and fitted values for the visibility $(V_d^c,V_d)$ for different dimensions $d$. The $1\sigma$ standard deviations are based on Poisson statistics.}
\begin{indented}
\item[]\begin{tabular}{@{}lllll}
\br
$d$  & $V_d^c$& $V_d$&\\\mr
2&0.707&0.903$\pm$0.018\\
3&0.775&0.860$\pm$0.019\\
4&0.817&0.959$\pm$0.008\\
\br
\end{tabular}
\end{indented}
\end{table}

\subsection{Time-bin basis}
\begin{figure}[ht]
\begin{center}
\includegraphics[width=0.8\textwidth]{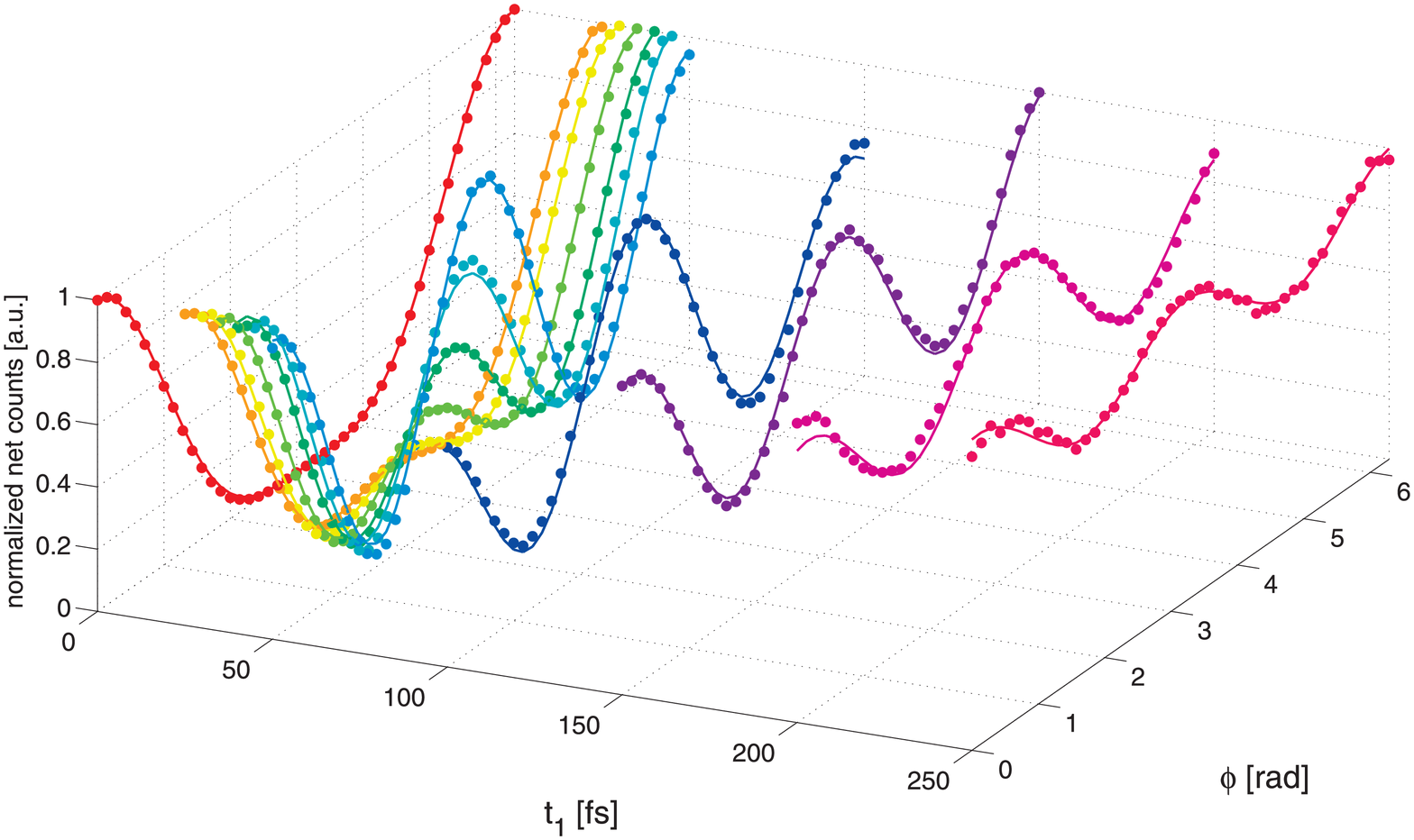} 
\caption{\label{fig:time_bin_qudits} Time-bin basis: Normalized two-photon interferences for a qubit encoded in time bins. The width of the bins is chosen to be $\Delta t_0=\Delta t_1=0$~fs and $t_0$ is fixed to $t_0=0$~fs. Shown are background-subtracted and normalized coincidence counts (normalized net counts). For $t_1=0$~fs the experimental data are fitted with (\ref{eq:signal_single}) (solid line). Since the visibility of the measured signal is equal to one no additional fitting parameter is needed. All measurements for $t_1>0$~fs are fitted with (\ref{eq:gamma_1gamma_2}) (solid lines) which involves the fitting parameters $\gamma_1$ and $\gamma_2$ depicted in the inset of figure~\ref{fig:I2}.}  
\end{center}
\end{figure}

\begin{figure}[ht]
\begin{center}
\includegraphics[width=1\textwidth]{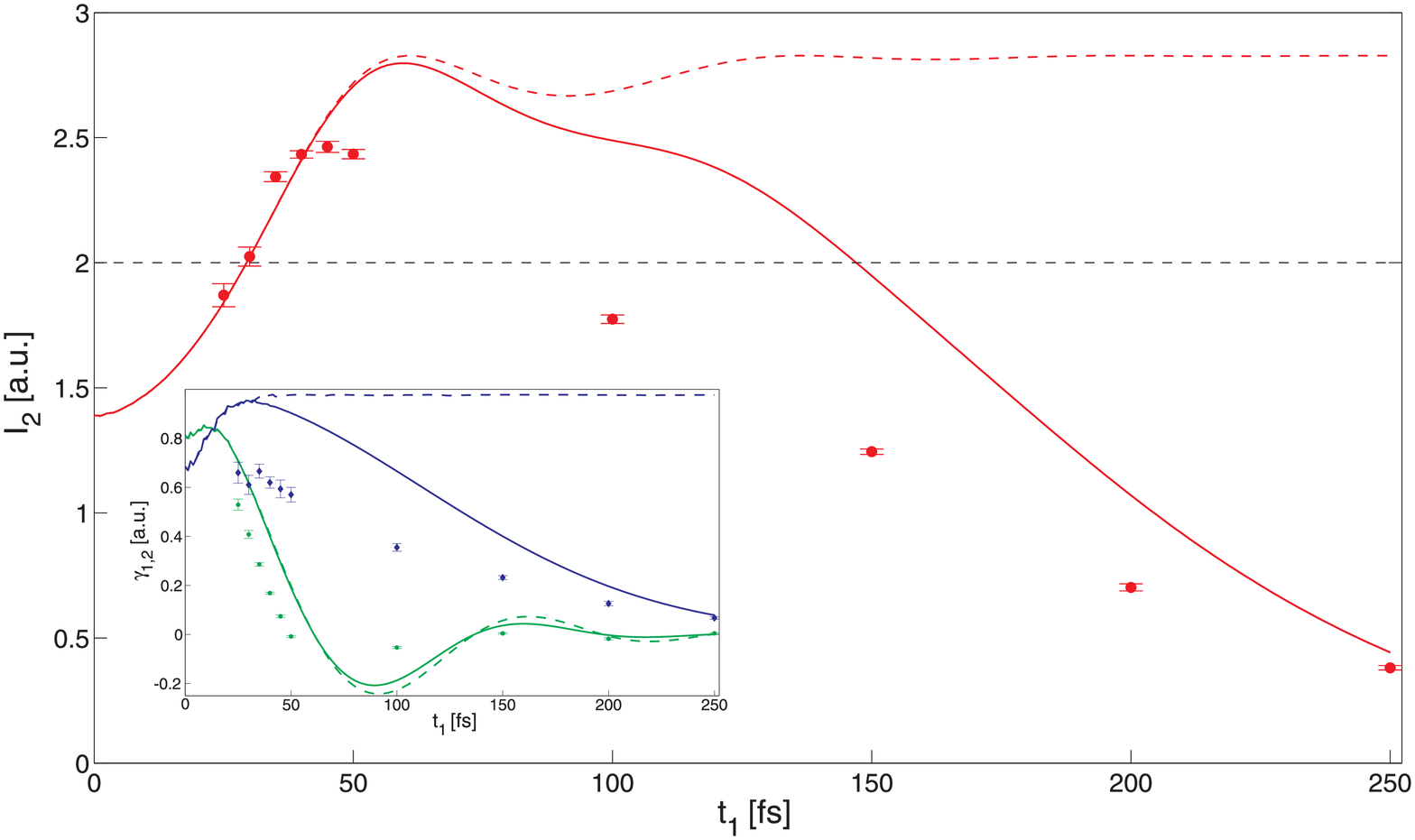} 
\caption{\label{fig:I2} The red dots show the measured Bell parameter $I_2$ in dependence of the experimentally evaluated fitting parameters $\gamma_1$ and $\gamma_2$ for various $t_1$ with 1$\sigma$ uncertainties. The dashed black line indicates the local realism limit. Theoretical predictions are calculated using the expression for $S$ of (\ref{eq:signal}) together with (\ref{eq:gamma}) (red dashed) and (\ref{eq:gamma_PSF}) (red solid) taking into account the finite spectral resolution of the experimental setup. Inset: The fitting parameters $\gamma_1$ (green dots) and $\gamma_2$ (blue diamonds) of (\ref{eq:gamma_1gamma_2}) obtained by the measurements in figure~\ref{fig:time_bin_qudits} as a function of $t_1$ with 1$\sigma$ errors.  Theoretical curves for $\gamma_1$ and $\gamma_2$ are calculated with (\ref{eq:signal}) using (\ref{eq:gamma}) ($\gamma_1$: green dashed, $\gamma_2$: blue dashed) and (\ref{eq:gamma_PSF}) ($\gamma_1$: green solid, $\gamma_2$: blue solid).}  
\end{center}
\end{figure}

According to (\ref{eq:time_bin_dt0}), a qubit is encoded in time bins via the applied transfer function
\begin{equation}\label{eq:M_time_bins_SLM}
M^{i,s}(\omega)=\vert u_0^{i,s}\vert+\vert u_1^{i,s}\vert\e^{i(\omega t_1+\phi_{i,s})}.
\end{equation}
Here, the time bin $f_0^{i,s}(t)$ is fixed at $t_0=0$~fs and bin $f_1^{i,s}(t)$ is delayed by $t_1$. To obtain maximally entangled states and fulfill condition (\ref{eq:M_condition}), we choose $\vert u_0^{i,s}\vert=\vert u_1^{i,s}\vert=1/2$. This is equivalent to use beam splitters with $T=R=1/2$ in the Franson experiment (figure~\ref{fig:time_bins_franson}). Although we set $\Delta t_0=\Delta t_1=0$ fs to make the individual bins as narrow as possible, the time bins, however, are always of finite width since the transfer function of (\ref{eq:M_time_bins_SLM}) is limited in $\omega$ due to the finite aperture of the SLM. Note, that in the present experiment the coherence time of the entangled photons is always larger than $\Delta t_j$ and is thus the limiting factor for the minimal $t_{1}$. Figure~\ref{fig:time_bin_qudits} depicts qubit interference traces for various positions $t_1$ of time bin $f_1^{i,s}(t)$. Again, the phase $\phi_i=\phi_s=\phi$ is varied by the SLM.
%\begin{figure}[ht]
%\begin{center}
%\includegraphics[width=1\textwidth]{gamma1_gamma2_theory.eps} 
%\caption{\label{fig:gamma} The fitting parameters $\gamma_1$ (green dots) and $\gamma_2$ (blue diamonds) of (\ref{eq:gamma_1gamma_2}) obtained by the measurements in figure \ref{fig:time_bin_qudits} as a function of $t_1$. Errors are $1\sigma$. Theoretical curves for $\gamma_1$ and $\gamma_2$ are calculated with $\Gamma(\omega_i,\omega_s)$ ($\gamma_1$: green dashed, $\gamma_2$: blue dashed) and $\Gamma_{PSF}(\omega_i,\omega_s)$ ($\gamma_1$: green solid, $\gamma_2$: blue solid) using (\ref{eq:signal}).}  
%\end{center}
%\end{figure}
If both bins completely overlap, i.e.~$t_1=0$~fs, the coincidence signal obviously consists of a product of two single photon interference rates
\begin{eqnarray}\label{eq:signal_single}
S^{(2)}(\phi)&=&\left\vert\langle\chi\vert\psi\rangle^{(2)}\right\vert^2 \nonumber\\
&\propto & \left\vert 1+\e^{i(\phi+\phi_0/2)}\right\vert^2 \left\vert 1+\e^{i(\phi+\phi_0/2)}\right\vert^2\nonumber\\
&\propto & \cos^4\left(\frac{\phi+\phi_0/2}{2}\right)
\end{eqnarray}
with 
\begin{equation}\label{eq:stat_sep_state}
\vert\psi\rangle^{(2)} = \frac{1}{2}\left(\vert 0 \rangle_i +\e^{i\phi_0/2}\vert 1 \rangle_i \right)\left(\vert 0 \rangle_s +\e^{i\phi_0/2}\vert 1 \rangle_s \right)
\end{equation}
and $\vert\chi\rangle$ of (\ref{eq:chi_phi}). As in (\ref{eq:qudit_maxent_phase}), the phase shift $\phi_0$ takes into account all uncompensated group velocity dispersion in the experimental setup. Equation (\ref{eq:signal_single}) is used to fit the experimental data in figure~\ref{fig:time_bin_qudits} for $t_1=0$~fs. Between $t_1\approx 25$~fs and $t_1\approx 50$~fs the contribution to the coincidence rate due to single photon interference decreases since $t_1$ begins to exceed the coherence time of the entangled photons. Consequently, the signal in figure~\ref{fig:time_bin_qudits} approaches the interference pattern of a maximally entangled qubit. The transformation from a non-entangled to a maximally entangled state can only be measured because sum frequency generation in a nonlinear crystal offers a coincidence window on the same time scale as $\tau_{i,s}^{coh}$. Note, that all measurements in figure~\ref{fig:time_bin_qudits} are performed for $t_1\ll\tau^{coh}_p\approx 88$~ns. To model the transition from a non-entangled qubit to a maximally entangled qubit we consider the state
\begin{equation}\label{eq:stat_g1_g2}
\fl\vert\psi\rangle^{(2)} = \frac{1}{\sqrt{1+2\gamma_1^2+\gamma_2^2}}\left(\vert 0 \rangle_i \vert 0\rangle_s+\gamma_1\e^{i\phi_0/2}\left[\vert 0 \rangle_i \vert 1\rangle_s+\vert 1 \rangle_i \vert 0\rangle_s\right]+\gamma_2\e^{i\phi_0}\vert 1 \rangle_i \vert 1\rangle_s\right)
\end{equation}
with additional parameters $\gamma_1$ and $\gamma_2$ quantifying the contribution of the one-photon and two-photon interferences to the coincidence rate 
\begin{eqnarray}\label{eq:gamma_1gamma_2}
S^{(2)}_{\gamma_1,\gamma_2}(\phi)&=&\left\vert\langle\chi\vert\psi\rangle^{(2)}\right\vert^2 \nonumber\\
&\propto &\left\vert 1+2\gamma_1\e^{i(\phi+\phi_0/2)}+\gamma_2 \e^{i(2\phi+\phi_0)}\right\vert^2.
\end{eqnarray}
Equation (\ref{eq:gamma_1gamma_2}) serves to fit the data points in figure~\ref{fig:time_bin_qudits} for $t_1>0$~fs where the obtained values for $\gamma_1$ and $\gamma_2$ are depicted in figure~\ref{fig:I2} (inset). The signal of (\ref{eq:gamma_1gamma_2}) has the property that 
\begin{equation}\label{eq:limes_gamma}
\lim_{\gamma_1 \to 0}S^{(2)}_{\gamma_1,\gamma_2}(\phi)=S_\lambda^{(2)}(\phi)
\end{equation}
with $S_\lambda^{(2)}(\phi)$ of (\ref{eq:fitd2}) and  $\lambda_2=V_2=2\gamma_2/(1+\gamma_2^2)$. Figure~\ref{fig:time_bin_qudits} demonstrates a decreasing visibility for large values of $t_1$. The realization of a qutrit, however, involves an additional time bin $f_2^{i,s}(t)$ centred at $t_2$ where $\Delta t_{21}>\tau_{i,s}^{coh}$. To obtain a qutrit with high visibility therefore requires an improved set up as discussed in section \ref{sec:conclusion}. 

For maximally entangled states, the CGLMP inequality can be related to the visibility $V_2$ of the interference fringes as discussed in section \ref{sec:cglmp}. On the other hand, the visibility is not a well-defined quantity in the coincidence signal based on the state of (\ref{eq:stat_g1_g2}) with $\gamma_1>0$. To study whether the generated qubits reveal entanglement we thus consider the Bell parameter $I_2$. This parameter is commonly determined by a series of projective measurements onto (\ref{eq:chi_phi}) for specific angles $(\phi_i,\phi_s)$ which constitute $I_2$ \cite{collins2002}. For the set of $(\phi_i,\phi_s)$ provided in \cite{collins2002}, we compute $I_2$ by a combination of single projection signals expressed as   
\begin{equation}\label{eq:s_bell}
S^{(2)}_{\gamma_1,\gamma_2}(\phi_i,\phi_s)
\propto\left\vert 1+\gamma_1 \e^{i(\phi_i+\phi_s)}+\gamma_2(\e^{i\phi_i}+\e^{i\phi_s})\right\vert^2
\end{equation}
and the experimentally fitted values for $\gamma_1$ and $\gamma_2$ shown in figure~\ref{fig:I2} (inset). Note, that the corresponding values for $I_2$ (figure~\ref{fig:I2}) are in fact underestimated since the settings for $(\phi_i,\phi_s)$ used to calculate the Bell parameter are only optimal in the case of maximally entangled qubits i.e.~for $\gamma_1=0$ and $\gamma_2=1$. The theoretical curves for $I_2$ are based on (\ref{eq:signal}) with $\Gamma(\omega_i,\omega_s)$ (red dashed) and $\Gamma_{PSF}(\omega_i,\omega_s)$ (red solid) where all experimental parameters used in the simulations are determined by other measurements \cite{bernhard2013thesis}.
The result which involves $\Gamma_{PSF}(\omega_i,\omega_s)$ shows decreasing values of $I_2$ for increasing $t_1$ due to the finite spectral resolution at the SLM. A similar behaviour can be observed in the measurement (red dots). Since the states incorporated to calculate the theoretical curves are pure, the remaining deviation between theory and data points is therefore caused by impurities in the experimentally realized state in combination with imperfections in the alignment of the optical setup which are not included in the theoretical model. The experimental data in figure~\ref{fig:I2} show a Bell parameter $I_2>2$, and thus entanglement, for $t_1$ between 35~fs and 50~fs. 
\subsection{Schmidt mode basis}
\begin{figure}[ht]
\begin{center}
\includegraphics[width=1\textwidth]{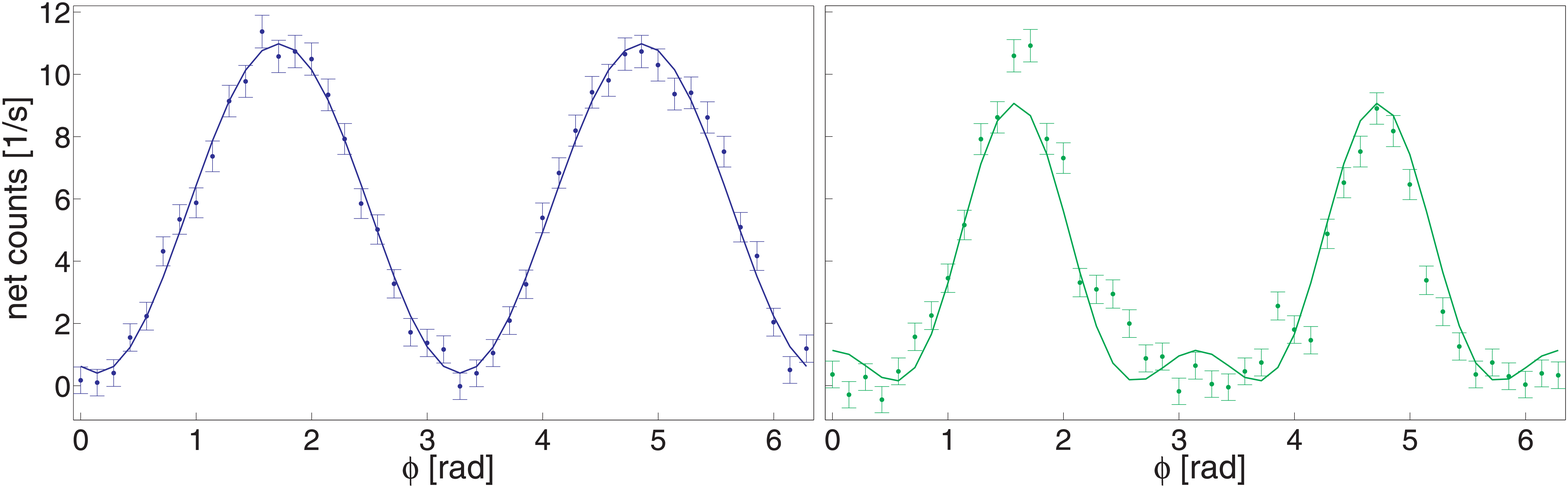} 
\caption{\label{fig:schmidt_qudits}Schmidt mode basis: Two-photon interferences for an entangled qubit (blue, left) and qutrit (green, right). Shown are background-subtracted coincidence counts (net counts) with 1$\sigma$ standard deviations using Poisson statistics. The solid curves are fits to the data points by means of (\ref{eq:fitd2}) and (\ref{eq:fitd3}).}  
\end{center}
\end{figure}

To encode qudits in the Schmidt decomposition of the idler and signal photon, the transfer function (\ref{eq:mslm}) consists of a linear combination of the calculated Schmidt modes $f_j^{i,s}(\omega)$ for our two-photon state. Figure~\ref{fig:schmidt_qudits} shows the corresponding two-photon interference fringes where $\phi_j^{i,s}=j\phi$ is the phase parameter between the modes according to (\ref{eq:mslm}). Equations (\ref{eq:fitd2}) and (\ref{eq:fitd3}) are used to fit the data points. Due to the combination of even and odd Schmidt modes an additional phase shift of $\pi/2$ can be observed in the measured curves. The corresponding fitting parameters for the visibility are summarized in table~\ref{tab:schmidt_deco_qubit}. In both measurements, the critical value $V_d^c$ for a Bell violation is exceeded and thus entanglement is present.
\begin{table}
\caption{\label{tab:schmidt_deco_qubit}Schmidt mode basis. Critical and fitted values of the visibility $(V_d^c,V_d)$. The $1\sigma$ standard deviations are based on Poisson statistics.}
\begin{indented}
\item[]\begin{tabular}{@{}lllll}
\br
$d$  & $V_d^c$& $V_d$&\\\mr
2&0.707&0.929$\pm$0.040\\
3&0.775&0.969$\pm$0.095\\
\br
\end{tabular}
\end{indented}
\end{table}

\subsection{Experimental limitations}
Using a frequency-bin discretization, we have demonstrated maximally entangled qudits up to $d=4$. The actual limitation to reach higher dimensions is the finite spectral resolution at the position of the SLM. If we increase the density of bins, frequencies from adjacent bins will overlap and the orthonormality condition of (\ref{eq:orthonormality}) begins to fail. This leads to a decrease in the entanglement due to non-vanishing $\vert 0 \rangle_i \vert 1\rangle_s $, $\vert 1 \rangle_i \vert 0\rangle_s$ contributions in (\ref{eq:psidisc}). For time bins, the reduction of the Bell parameter for large time delays between the two bins can also be attributed to the finite spectral resolution. Here, it currently constrains even more the maximally accessible dimension. Both of these methods have, however, the advantage that a perfect knowledge of the two-photon state is not required in order to discretize a maximally entangled state. In contrast, the Schmidt decomposition is very sensitive to the form of the state generated by SPDC. It is likely, that the theoretical $\Gamma_{PSF}(\omega_i,\omega_s)$ deviates from the joint spectral amplitude realized in the experiment due to a lack in the precise knowledge of all the physical parameters used to compute the Schmidt modes. This lowers the quality of the measured two-photon interference patterns in terms of their shape and the obtained coincidence count rate.  

\section{Conclusion and outlook}\label{sec:conclusion}
We have presented a thorough theoretical description how to encode qudits in the frequency domain of energy-time entangled photons whose coincidences are detected through SFG in a nonlinear crystal. Although applied to the specific case of a SLM, the discretization procedure of the frequency space is general and describes a unified framework for different energy-time resolving experimental schemes. It has been discussed, that the entanglement content in a two-photon state generated by continuous wave parametric down-conversion is very high. However, it is reduced if the finite spectral resolution of the experimental setup is accounted for. The flexibility of a SLM has been exploited in order to project the energy space of the entangled photons onto different bases. In particular, we implemented frequency and time bins to measure maximally entangled qudits through two-photon interference fringes. All qudits have been investigated in view of their entanglement properties by means of a generalized Bell inequality. The time-bin scheme allowed to show the transition from a separable to a maximally entangled qubit state taking advantage of an ultrafast detection method with femtosecond temporal resolution. In addition, we expressed the two-photon wave function in a Schmidt decomposition to use the resulting modes as a further basis for qudits. 

As the current limitation in the quality and dimension of the generated states, we identified the finite spectral resolution at the position of the SLM in combination with the size and number of its pixels. A way to improve the spectral resolution is to replace the prisms with gratings. This would allow to spatially disperse the spectrum along a wider range of the SLM display. The achievable dimension becomes then only limited by the number of pixels of the shaping device. The low efficiency coincidence detection method using SFG therefore constitutes a further bound on the dimension of the qudits. A higher coincidence rate could be achieved by generating SFG in a waveguide instead of a bulk crystal \cite{sangouard2011} or using enhanced detection schemes \cite{sensarn2009}. Ultimately, time synchronized sum-frequency generation between a SPDC photon and a femtosecond laser pulse provides an ultrafast coincidence detection method with an upconversion efficiency per pump pulse of about 25\% as demonstrated in \cite{kuzucu2008}. A similar scheme could be realized using two spatially separated upconversion crystals allowing to perform an ultrafast and non-local coincidence detection on separated photons. This, however, additionally requires the replacement of the continuous wave pump with a pulsed pump laser in order to increase the duty cycle. Such a modification of the here presented experimental setup would finally provide the necessary detection efficiency to perform quantum communication processing.     

That a state can be precisely characterized in terms of its Schmidt decomposition was shown in \cite{straupe2011} for transverse momentum entangled photons. The here presented experimental setup allows for projective measurements on single Schmidt modes $f_j^{i,s}(\omega)$ in the frequency domain. A state reconstruction similar to \cite{straupe2011} can then be performed in the entangled photons energy-time degrees of freedom provided the correct eigenfunctions are known. The obtained $\beta_j$ can further be used to calculate the Schmidt number to quantify the entanglement in the state. 

The above mentioned improvements will allow to encode qudits in dimensions inaccessible by standard interferometry. High-dimensional entangled quantum states could find applications in future quantum communication systems where energy-time entangled photons play a key role \cite{gisin2007}. The method presented here allows to manipulate and characterize this entanglement in a very flexible way.

\section{Acknowledgements}
This research was supported by the grant PP00P2\_133596 and by the NCCR MUST, both funded by the Swiss National Science Foundation.

\end{document}